\documentclass[11pt]{article}
\usepackage{a4wide}
\usepackage{amsmath,amssymb,amsfonts}
\usepackage{comment}
\usepackage{epsfig,verbatim,graphics}
\usepackage{subfigure,slashed}

\usepackage{cleveref}
\crefformat{pluralequation}{#2\black{eqs.~(}#1\black{)}#3}
\Crefformat{pluralequation}{#2\black{Equations~(}#1\black{)}#3}
\crefformat{pluralfigure}{#2\black{figs.~}#1#3}
\Crefformat{pluralfigure}{#2\black{Figures~}#1#3}

\newcommand{\mathsym}[1]{{}}

\newcommand{\eref}[1]{eq.~(\ref{#1})}

\renewcommand\({\left(}
\renewcommand\){\right)}
\renewcommand\[{\left[}
\renewcommand\]{\right]}

\newcommand{\dd}{{\rm d}}
\declareslashed{}{{^-}}{.1}{0.1}{\rm d}

\newcommand{\e}{{\rm e}}

\newcommand\eps{\epsilon}
\newcommand\mpl{m_{\rm p}}

\def\be{\begin{equation}}
\def\ee{\end{equation}}
\def\ba{\begin{eqnarray}}
\def\ea{\end{eqnarray}}

\def\O{\mathcal{O}}

\def\nn{\nonumber}
\def\({\left(}
\def\){\right)}

\usepackage{tikz}
\usetikzlibrary{arrows,shapes}
\usetikzlibrary{trees}
\usetikzlibrary{matrix,arrows} 				
\usetikzlibrary{positioning}				
\usetikzlibrary{calc,through}				
\usetikzlibrary{decorations.pathreplacing}  
\usepackage{pgffor}							

\usetikzlibrary{decorations.pathmorphing}	
\usetikzlibrary{decorations.markings}
\tikzset{
    vector/.style={decorate, decoration={snake}, draw},
	provector/.style={decorate, decoration={snake,amplitude=2.5pt}, draw},
	antivector/.style={decorate, decoration={snake,amplitude=-2.5pt}, draw},
    fermion/.style={draw=black, postaction={decorate},
        decoration={markings,mark=at position .55 with {\arrow[draw=black]{>}}}},
    fermionbar/.style={draw=black, postaction={decorate},
        decoration={markings,mark=at position .55 with {\arrow[draw=black]{<}}}},
    fermionnoarrow/.style={draw=black},
    gluon/.style={decorate, draw=black,
        decoration={coil,amplitude=4pt, segment length=5pt}},
    scalar/.style={dashed,draw=black, postaction={decorate},
        decoration={markings,mark=at position .55 with {\arrow[draw=black]{>}}}},
    scalarbar/.style={dashed,draw=black, postaction={decorate},
        decoration={markings,mark=at position .55 with {\arrow[draw=black]{<}}}},
    scalarnoarrow/.style={dashed,draw=black},
    electron/.style={draw=black, postaction={decorate},
        decoration={markings,mark=at position .55 with {\arrow[draw=black]{>}}}},
	bigvector/.style={decorate, decoration={snake,amplitude=4pt}, draw},
}

\newcommand{\roughly}[1]{\mathrel{\raise.3ex\hbox{$#1$\kern-0.85em
\lower1ex\hbox{$\sim$}}}}

\begin{document}

\begin{titlepage}
\begin{center}

\today

\vskip 1.5cm

{\LARGE \bf Disformal transformations as a change of units }\\[1cm]

\vskip 1cm

\renewcommand*{\thefootnote}{\fnsymbol{footnote}}
\setcounter{footnote}{0}

{\bf
    Jacopo Fumagalli$^{1}$\footnote{{\tt ja.fumagalli@gmail.com},
        \footnotemark[2]{\tt sander.mooij@epfl.ch},
        \footnotemark[3]{\tt mpostma@nikhef.nl}},
    Sander Mooij$^{2}$\footnotemark[2] and
    Marieke Postma$^1$\footnotemark[3]
}

\renewcommand*{\thefootnote}{\number{footnote}}
\setcounter{footnote}{0}

\vskip 25pt

{\em $^1$ \hskip -.1truecm
Nikhef, \\Science Park 105, \\1098 XG Amsterdam, The Netherlands
}

\vskip 20pt

{\em $^2$ \hskip -.1truecm
Institute of Physics, Laboratory of Particle Physics and Cosmology,\\ \'Ecole Polytechnique F\'ed\'erale de Lausanne,\\ CH-1015 Lausanne, Switzerland 
}

\end{center}

\vskip 0.5cm

\begin{center} {\bf ABSTRACT}\\[3ex]
\end{center}

\noindent A disformal transformation is a very useful tool to analyze a general effective theory description of inflation. It can for example be used to set the tensor sound speed to unity, such that the tensor power spectrum only depends on the Hubble parameter. However, the disformal transformation has also led to quite some confusion in recent literature. We hope to clarify that confusion by pointing out that a disformal transformation is nothing else than a change of units. We show how everything that can be achieved by a (possibly time-dependent) disformal transformation, equally follows from the equivalent change of units, up to all orders in perturbation theory. We also comment on the sensitivity of the tensor power spectrum to a non-standard tensor speed of sound.



%
%

\end{titlepage}

\newpage
\setcounter{page}{1}


\section{Introduction}

Gravitational waves, or tensor modes, are a generic prediction of inflation. In Einstein gravity, gravitational waves propagate at the speed of light. Therefore, the only unknown in the tensor power spectrum is the Hubble scale during inflation, and the detection of primordial gravitational waves from inflation would unambiguously pin down this scale.  This direct connection is lost in a more general theory of gravity, as now the tensor spectrum depends on both the Hubble scale and on the speed of sound of the tensor modes.

A model-independent approach to (single-field) inflation is provided by the effective field theory (EFT) of inflation \cite{EFT}.  The EFT allows for a non-canonical tensor speed $c_T \neq 1$.\footnote{We work in units $\hbar = c =1$.  Speeds are this given in units of the speed of light  {\it today}: $c = 3 \times 10^8$ m/s.} Physical motivations can be found in brane world models or models of modified gravity, such as Horndeski's most general scalar-tensor theory with second-order field equations \cite{Horndeski}.  It should be noted that although the tensor and light speed may differ during inflation, they should evolve to equal values today.  Last year's detection of gravitational waves from the binary neutron star merger GW170817, and the subsequent observation of the electromagnetic counterpart, gives the very stringent bound on the tensor speed today $|c_T-1| \leq 3 \cdot 10^{-15}$ \cite{Monitor:2017mdv}.

Recently, there has been a lot of interest in disformal transformations of the spacetime metric, called ``disformal" because the metric's temporal component is treated in a different way than its spatial component. In general one writes
\be \label{disformal0}
g_{\mu\nu} \mapsto \tilde  g_{\mu \nu} 
= \Omega^2(g_{\mu \nu} +(1-B^2)  n_\mu n_\nu),
\ee
where $B$ generates a pure disformal transformation and $\Omega$ a subsequent conformal transformation, that can be used to normalize the Planck mass.  In \cite{Creminelli}, and later in \cite{felice}, it was proposed to use such a disformal transformation --- with $\Omega^{-2} =B$ and $n^\mu$ the unit normal to surfaces of constant time \cite{EFT} --- to set any non-canonical tensor speed $c_T$ to unity (that is, to the light speed today).  In the new frame, the Einstein frame, gravity is of standard form. The tensor speed does not appear explicitly in the tensor power spectrum anymore. It thus follows immediately that all tensor observables, such as the power spectrum and the bispectrum, can only depend on the Hubble scale defined in this frame \cite{Creminelli}.

However, it is important to realize that this does not mean that in the Einstein frame all dependence on the tensor speed has dropped out of the model.  For example, in the background Friedmann equation which relates the Hubble constant to the microscopic model of inflation now a factor of $c_T$ appears. The disformal transformation does not only rescale the tensor speed, but the sound speeds of all other species in the universe as well.  In fact the ratio of the tensor to light speed during inflation remains invariant $c_T/c_A = \tilde{c}_T/\tilde{c}_A$.  This is in agreement with \cite{Misao2} who showed that a pure disformal transformation ($\Omega=1$) cannot change the causal structure and the propagation speed of the fields, since it is equivalent to a redefinition of the time coordinate.  See also the discussion in \cite{Ellis, Ellis2}, in which it is pointed out that only changing the time coordinate of the metric does not change the physical speed of light.  Thus although the tensor propagator, and therefore tensor quantum loop corrections, are standard in the Einstein frame, the propagator equation and loop corrections of all other fields will generically be non-standard (and depend on $c_T$).

In this paper we point out that the disformal transformation can equally be seen as a change of units.  The mathematics of a disformal transformation and a change of units is one-to-one. Thinking in terms of a change of units is useful as it provides a physical and more intuitive understanding of what is going on. We will show how it helps to clarify some confusion in recently published literature.

This paper is organized as follows. After introducing the general EFT setup allowing for a non-canonical tensor speed in  \cref{setup}, and reviewing the resulting scalar and tensor power spectra in \cref{spectra}, we introduce the disformal transformation in \cref{disfor}. In \cref{uchange} we show that the disformal transformation can equally be viewed as a change of units.  In \cref{timdep} we discuss the time-dependent generalization of the disformal transformation, and \cref{bey} shows that the equivalence between the disformal transformation and a change of units does not only hold for the free theory (and for the power spectrum), but at every order in perturbation theory (and thus for the bispectrum and higher order correlation functions as well). We comment on the ``resilience" of the tensor spectrum, discussed in \cite{Creminelli}, in \cref{resil}, followed by a further discussion of results in recent literature in \cref{lit}.  This section shows how interpretating the disformal transformation as a change of units can make the physics more transparent. Finally, we conclude in \cref{con}.

\section{Set-up} \label{setup}

We are interested in the effect of a disformal transformation on the inflationary description.  The metric is that of a perturbed FRW universe, which in ADM decomposition reads \cite{ADM1,ADM2}
\be
\dd s^2 = -N^2  \dd t^2 + h_{ij}(N^i  \dd t + \dd x^i) (N^j  \dd t + \dd
x^j),
\label{metric1}
\ee
with scalar and tensor perturbations
\be
h_{ij} = a^2 \e^{2\zeta} 
(\e^{\gamma})_{ij},\qquad
\gamma_{ii} =0=\partial_i \gamma_{ij}.
\label{metric2}
\ee
The effective field theory (EFT) action for single field inflation in
unitary gauge is \cite{EFT}
\begin{align}
 S =&  \frac{ \mpl^2 }{2}\int \dd t \dd^3 \vec x \sqrt{-g} \bigg[
R - \rho(t) g^{00}-\Lambda(t)-(1-  c_T^{-2}) (\delta K_{\mu\nu}^2 -\delta K^2)
+ M_2^4 (\delta g^{00})^2
\bigg].
\label{action}
\end{align}
Here $\delta K_{\mu\nu}$ stands for the perturbation of $K_{\mu\nu}$, the extrinsic curvature on spatial slices
\be
K_{\mu\nu} = \frac{1}{2N} \left(\dot{h}_{ij} -\nabla_i N_j - \nabla_j N_i\right).
\ee
The last two terms in \cref{action} affect the speed of sound of tensor and scalar modes respectively; if non-zero the respective sound speeds will differ from unity. The scalar speed of sound is related to $M_2$ via $ c_s^2 -1=- {2 M_2^4} / ( \dot H^2 \mpl^2)$.

For simplicity, we focus on the quadratic action, but results can be generalized to the full theory as discussed in \cref{bey}. The free action for the tensor and scalar modes is \cite{EFT, Creminelli}
\begin{align}
S^{(2)}_{\gamma} &= \frac{\mpl^2}{8} \int \dd t \dd^3 x~ a^3~c_T^{-2} \[
\dot \gamma_{ij}^2 - c_T^2 \frac{(\partial_k \gamma_{ij})^2}{a^2} \],
\nn \\
 S^{(2)}_{\zeta}  &= \mpl^2  \int \dd t \dd^3 x~ a^3~ \eps~  c_s^{-2} \[
\dot \zeta^2 - c_s^2 \frac{(\partial_k \zeta)^2}{a^2} \] ,
\label{sound1}
\end{align}
with $\eps = -\dot H/H^2$ the first slow roll parameter. 


\subsection{The power spectrum} \label{spectra}

The tensor power spectrum follows from the quadratic action \cref{sound1}. In the superhorizon limit ${c_s k}/{(aH)}\ll1$ and up to slow-roll corrections, it is
\be
\Delta_\gamma^2 = \frac{2}{ \pi^2 \mpl^2}\frac{H^2}{c_T}
\left( \frac{c_T k}{aH} \right)^{-2\eps-\eps_T},
\label{Pgamma}
\ee
where we have defined the slow-roll parameter $\eps_T \equiv {\dot{c}_T}/{(H c_T)}$. 
For the tensor spectral index this directly gives  $n_T -1 
= -2\eps -\eps_T$. In the superhorizon limit, the tensor power spectrum is time independent.

A well-known, crucial observation is that in a theory with $c_T=1$, measuring the tensor amplitude directly gives the value of $H$ during inflation.  In addition,  a blue-tilted spectrum  ($n_T>1 $) automatically implies $\eps < 0 \, \Rightarrow \, \dot{H}>0$, that is,  a violation of the Null Energy Condition (NEC).  Clearly, these two robust predictions no longer hold for a non-canonical tensor speed $c_T \neq 1$, which arises if the term proportional to $(\delta K_{\mu\nu}^2 - \delta K^2)$ in the inflationary EFT in \cref{action} is non-zero.

Reference \cite{Creminelli} has argued that without loss of generality, one can always perform a disformal transformation that sets $c_T=1$. The statement is that the tensor power spectrum is ``resilient": the tensor power spectrum cannot be modified as $c_T$ can be set to unity, and thus both the relation between the observed tensor amplitude and the scale of inflation, and the relation between a blue-tilted spectrum and NEC violation, are robust.  We will come back to this in section \cref{resil}.

The scalar power spectrum that follows from \cref{sound1} is
\be
\Delta_{\zeta}^2= \frac{1}{8 \pi^2\mpl^2}\frac{H^2}{c_s \eps}
\left( \frac{c_s k}{aH} \right)^{-2\eps-\eta-\eps_s},
\label{Pzeta}
\ee
which involves the additional slow-roll parameters $\eta \equiv {\dot{\eps}}/{(H \eps)}$ and $\eps_s \equiv  {\dot{c}_s}/{(H c_s)}$. The scalar spectral index is defined as $n_s-1 
= -2\eps -\eta-\eps_s$.

\section{Disformal transformation} \label{disfor}

Consider the combined conformal (described by $\Omega$) and disformal
(described by $B$) transformation
%
\be
g_{\mu\nu} \mapsto \tilde  g_{\mu \nu} = \Omega^2 (g_{\mu \nu} +(1-B^2)  n_\mu n_\nu),
\label{disformal}
\ee
with $n^\mu$ the unit normal to surfaces of constant $t$: \cite{EFT}
\be
n_\mu =\frac{\partial_\mu t}{\sqrt{-g^{\mu\nu} \partial_\mu t \partial_\nu t}} = N \delta_{\mu 0}.
\ee
In terms of the metric components the transformation reads
\ba
g_{00}\mapsto  \tilde{g}_{00} &=&
\Omega^2\left(-B^2N^2 +h_{ij} N^i N^j \right)\nn\\
g_{0i} \mapsto  \tilde{g}_{0i} &=&
\Omega^2 h_{ij} N^j\nn\\
g_{ij} \mapsto  \tilde{g}_{ij} &=& \Omega^2 g_{ij}=\Omega^2 h_{ij},
\ea
and the line element transforms as
\be
\dd s^2 \mapsto \dd\tilde{s}^2= -\Omega^2 B^2 N^2 d t^2 +\Omega^2  h_{ij}(  N^i d t + d x^i) ( N^jd t + dx^j).   \label{resul}
\ee
The combined con/disformal transformation -- from now on sloppily referred to as ``disformal transformation"
--  is equivalent to a field/coordinate redefinition that satisfies
\be
\tilde N \dd \tilde t =\Omega B N \dd t, \quad
\tilde h_{ij}(\tilde t) \dd \tilde x^i \dd \tilde x^j
=\Omega^2  h_{ij}(t) \dd x^i \dd x^j ,\quad
\tilde h_{ij}(\tilde t) \tilde N^i \tilde N^j \dd \tilde t \dd \tilde t
=\Omega^2  h_{ij}(t) N^i N^j\dd t \dd t .
\label{tilde}
\ee
One can choose to keep the coordinates invariant $\dd \tilde x = \dd x$ and $\dd t = \dd \tilde t$, then the transformation is a pure field redefinition.  Here we follow the choice made in \cite{Creminelli} to keep $\dd x$ and $N$ invariant instead.  It can be checked explicitly that  applying the disformal transformation \cref{disformal} to the EFT action \cref{action} is equivalent to the field redefinition \cref{tilde}.\footnote{The EFT for inflation \cref{action} is not written in covariant form.  However, if its UV completion is a covariant theory, it follows that $\Lambda$ behaves as a scalar, and $\rho$ and $M_2$ as $T_{00}$-tensors under a disformal transformation.}

The choice $\Omega^2 = B^{-1}=\beta$ will change the speed of sound while keeping the normalization of the Planck mass fixed.  The field redefinition \cref{tilde} then amounts to
\begin{align}
  \tilde{dt} = \frac{1}{\sqrt{\beta}} dt, \qquad
 \tilde{a} = \sqrt{\beta} a, \quad
\tilde{N^i} = \sqrt{\beta} N^i,
          \label{betatra}
\end{align}
and $\dd \tilde x = \dd x$ and $\tilde N = N$ invariant.  Physically, performing a disformal transformation comes down to stretching /squeezing time and space intervals. Time intervals pick up a factor $\beta^{-1/2}$, space intervals pick up a factor $\beta^{1/2}$ (physical distances are given by $a \dd x$). The effect on any velocity $c_X$ is thus given by
\be
c_X \mapsto \tilde{c}_X = \frac{\tilde{a} \tilde{\dd x}}{\tilde{\dd t}}  
=\beta \frac{a \dd x}{ \dd t}
= \beta c_X.  \label{trafc}
\ee
The change in velocity can equivalently be derived from the quadratic action \cref{sound1}, which after a disformal transformation becomes 
\begin{align}
S_{\gamma} &=\frac{\mpl^2}{8} \int \dd \tilde t \dd^3 \tilde x
  \tilde a^3 
 \frac{1}{\tilde c_T^2} \[
\(\partial_{\tilde t} \gamma_{ij}\)^2 - \tilde c_T^2
  \frac{(\partial_k \gamma_{ij})^2}{ \tilde a^2} \],
\nn \\
S_{\zeta} &=\mpl^2 \int \dd \tilde t \dd^3 \tilde x
  \tilde a^3 \frac{1}{\tilde c_s^2} \eps \[
\(\partial_{\tilde t} \zeta\)^2 - \tilde c_s^2
  \frac{(\partial_k \zeta)^2}{ \tilde a^2} \],
\label{sound2}
\end{align}
where $\tilde c_T = \beta c_T$ and $\tilde c_s = \beta c_S$ can indeed be identified as the sound speeds in the new frame.

We close this section by taking a look at the tensor and scalar power spectrum expressed in the tilde variables of the new frame:
\begin{align}
  \Delta_\gamma^2 &=  \frac{2}{ \pi^2 \mpl^2}\frac{\tilde{H}^2}{\tilde{c}_T} \left( \frac{\tilde{c}_T \tilde{k}}{\tilde{a}\tilde{H}} \right)^{-2\tilde{\eps}-\tilde{\eps}_T} , \nn \\
\Delta_{\zeta}^2&= \frac{1}{8 \pi^2\mpl^2}\frac{\tilde H^2}{\tilde c_s (\tilde\eps + \frac{\beta_{\tilde t}}{2})}
                  \left( \frac{\tilde c_s k}{\tilde a \tilde H} \right)^{-2\tilde \eps-\tilde \eta(1+...)-\tilde \eps_s},
\label{tildeP}                  
\end{align}
where the ellipses in the exponent of the scalar power spectrum are  $\O( \frac{\beta_{\tilde t}}{\tilde \eps},\,\beta_{\tilde t \tilde t})$ corrections, with $\beta_{\tilde t} = \frac{\partial_{\tilde t} \beta}{\tilde H \beta}$ and
 $\beta_{\tilde t \tilde t } = \frac{\partial_{\tilde t} \beta_{\tilde t}}{\tilde H \beta}$.
 The functional form of the tensor power spectrum is invariant under a disformal transformation, as can be seen comparing \cref{Pgamma,tildeP}.  This is, however, not the case for the scalar power spectrum which has a different functional form in the new frame as can be seen comparing with \cref{Pzeta}.  This can be traced back to the quadratic scalar action \cref{sound2} which depends on $\eps = \tilde \eps + \frac{\beta_{\tilde t}}{2}$.  Of course, physical observables are frame-independent, and the numerical value of the power spectrum is the same in both frames.  The invariance of the power spectrum has been discussed in the literature \cite{Creminelli, felice,Misao2,Minamitsuji,Tsujikawa:2014uza,Tsujikawa,Misao1,  Arroja, Domenech:2015tca, Baumann, Cai, Cai2, Sebastian, Motohashi:2015pra,Chen}.  A particular feature may in one frame show up in the tensor speed, and in another frame in the Hubble parameter, but the resulting power spectrum always comes out equally.

Finally, we note that choosing $\beta = c_T^{-1}$ sets the tensor speed of sound in the new frame to unity $\tilde c_T =1$, and the tensor power spectrum is only a function of $\tilde H$.

\section{Disformal transformation as change of units}  \label{uchange}

After this short review, we are ready to deliver the core message of this paper: the disformal transformation is nothing else than a change of units. That is, the mathematics and all the results that follow from a disformal transformation can equally be intrepreted as a change of units. Admittedly, this is a relatively small step from the observation that the disformal transformation changes time and space intervals, as noted below \cref{betatra}. However, we think that viewing the abstract disformal transformation as a mere change of units does provide some very useful physical intuition, which given the many frame dependent statements and confusing statements in the literature is sometimes missing.

Let's make the equivalence even more concrete.  Introduce a length unit $L$ and time unit $T$ that measures length and time intervals in the original frame. In natural units they can be identified with the Planck length $l_{\rm p} = \sqrt{{\hbar G}/{c^3}}$ and Planck time $t_{\rm p} = \sqrt{\hbar G/{c^5}}$ respectively, but the story does not change if we choose to work with, say, meters and seconds instead.  Consider a finite length interval $\Delta x = \bar x \, L$ that measures $\bar{x}$ unit lenghts $L$, with $\bar{x}$ a dimensionless number. Now, let us take for simplicity a time-independent disformal transformation\footnote{For the time-dependent case the same argument holds using infinitesimal displacements (see next section).} that sends this length interval of $\bar x$ length units to $\beta^{1/2} \bar x $ units.\footnote{Note that physical length intervals are given by $\Delta x =\int a \dd x$, and the rescaling follows from the rescaling of the scale factor in \cref{betatra}.}  Equivalently, this can be viewed as the transformed length interval still measuring $\bar x$ length units, but with rescaled units $\tilde L = \beta^{1/2} L$.  Similar for finite time intervals $\Delta t = \bar t \, T$: the disformal transformation sends this time interval of $\bar t$ time units to $\beta^{-1/2} \bar t$ units, or equivalently to $\bar t$ rescaled time units $\tilde T = \beta^{-1/2}T$.  In formulas, the equivalence reads
\begin{align}
\Delta x = \bar x \, L  & \quad\mapsto \quad \Delta \tilde x = \bar x \beta^{1/2}  \, L = \bar x \, \tilde L, \nn \\
  \Delta t = \bar t \,  T & \quad\mapsto \quad\Delta \tilde t =\bar t \beta^{-1/2}\,  T = \bar t \, \tilde T.
                               \label{equal}
\end{align}
Since $(\tilde L/\tilde T) = \beta (L/T)$ the change of units changes all sound speeds $c_i \mapsto \tilde c_i =\beta c_i$. 

We focussed here on the special choice $\Omega^2 = B^{-1} = \beta$, but it is clear that the arguments can be generalized straightforwardly to the more general disformal transformation  \eref{disformal}.  It thus also follows that the purely conformal transformation ($B=1$), which can be used to transform a Lagrangian with a non-minimal coupling to one with Einstein-Hilbert gravity, is nothing but a change of units.  This was also noted in e.g. \cite{catena, volpe} who interpreted the conformal transformation as a rescaling of the Planck mass (in the natural unit system).

\subsection{Time-dependent disformal transformation}  \label{timdep}

Special care should be taken to implement a change of units (to perform a disformal transformation) if the transformation is time-dependent, i.e. $\beta = \beta(t)$.  Indeed, if the initial units $L,T$ are taken time-independent, the new units $\tilde L, \tilde T$ are time-dependent.  If a rate is expressed in the new unit system, for example the Hubble rate, there is a contribution because the quantity itself is time-dependent (in the case of the Hubble constant the physical distance bet\-ween galaxies changes with time in an expanding universe), but also a contribution because the unit itself changes with time. A naive dimensional analysis is not enough to deduce the transformation law.

Infinitesimal length and time intervals transform under a disformal transformation / change of units as in \cref{equal}, i.e. $d\tilde t=\beta^{-1/2}d\bar t T=d\bar t \tilde{T}$.  From this we can derive the transformation of the time derivative of a length $\Delta x(t)= \int \bar dx \, L$.
\begin{align}
\frac{\partial }{\partial t}  ( \Delta x )\quad \mapsto \quad \frac{\partial}{\partial \tilde{t}} \,  (\Delta \tilde{x})
= \frac1{\tilde T}\frac{\partial}{  \partial \bar  t }  \, \left(\int d\bar{x}\, \tilde{L}\right)
= \beta^{1/2} \,\frac{1}{T} \frac{\partial}{ \partial \bar{ t}  } \,  \left(\int d\bar{x} \beta^{1/2} L\right) 
=  \beta\left[ \partial_t(\Delta x) + \frac{1}{2} \frac{{\partial_t \beta}}{\beta } \Delta x\right],
\end{align}
%
As before, the factors of $\beta$ can be either viewed as arising from the disformal transformation or from the change of units. Whatever can be achieved by a disformal transformation can equivalently be obtained from a change of (in this case time-dependent) units.  

The Hubble rate is given by the dimensionless derivative of a length scale $\partial_t \ln (\Delta x)$, and thus transforms as
\be
H \mapsto \tilde{H} = \beta^{1/2}H\left(1+\frac{1}{2}\frac{\dot{\beta}}{\beta H}\right).
\label{HubbleT}
\ee
This relation can be
used to verify that the power spectra are the same in both frames.

In \cref{invariant} we show how the action can be written in terms of dimensionless quantities, which are manifestly invariant under a change of units.

\subsection{Beyond quadratic order}  \label{bey}

For simplicity, much of this paper focusses on the quadratic action and the power spectra. We want to emphasize though that the equivalence between a disformal transformation and a change of units holds at all levels in perturbation theory. One may argue\footnote{As an anonymous referee did.} that a disformal transformation contains more information, as it informs about the strucure of the EFT in \cref{action}.  This is not correct: our statement is that \eref{betatra} {\it is} a change of units, which thus gives equivalent results.  One may worry that the disformal transformation does not only rescale length and time intervals, but also acts on the shift vector $N^i$.  However, once we integrate out the non-dynamical lapse $N$ and shift $N^i$ both methods yield the same action for $\zeta$. Rescaling a Lagrange multiplier does not change the action resulting from integrating out that Lagrange multiplier.\footnote{Working in longitudinal gauge, where the shift function $N^i$ vanishes, the disformal transformation is one-to-one to a change of units also at the level of the unconstrained action. This gauge was used in \cite{Misao2} to show that a purely disformal ($\Omega =1$) transformation is equivalent to a time rescaling.}
In fact, that has been done to obtain the quadratic actions \cref{sound1,sound2}.

The disformal transformation can be used to eliminate certain operators from the EFT, and thus can be used to construct a minimal set of EFT operators.  Indeed, with the transformation $c_T \mapsto 1$ the term depending on the extrinsic curvature in \cref{action} is set to zero. Of course, the same result can be obtained with a change of units, which thus contains the same information about the structure of the EFT. In this case, the result is arguably even more transparent.

The same can be seen at the level of the correlation functions.  As an example, take $M_2=0$ in \cref{action} such that scalar speed is canonical $c_s=1$. The scalar three-point interaction $\langle \zeta \zeta\zeta\rangle$ gives the bispectrum and level of non-Gaussianity. A direct computation \cite{Creminelli} yields $f_{\rm NL} = 1- c_T^2$. The main message in \cite{Creminelli} is that one can actually avoid doing this computation, by mapping it to a computation done many times in the literature already. Indeed a disformal transformation with $\beta = c_T^{-1}$ brings us to the Einstein frame with $\tilde{c_T}=1$ and $\tilde{c_s} = \frac{1}{c_T}$. This set-up has been studied at great length, and it is known that the $\langle \zeta \zeta\zeta\rangle$ correlator yields $f_{\rm NL} = 1- \tilde{c}_s^{-2}$. In this way, the disformal transformation can be used for computations involving higher order operators in the inflationary EFT.  It is clear though that changing units helps us in an equivalent way. In units that set $\tilde c_T=1$, the scalar speed is $\tilde c_s = {1}/{c_T}$, and we get the Einstein frame results. Now switching back to units $c_s=1$ this becomes again $f_{\rm NL} = 1- c_T^2$.

\section{Applications}

\subsection{Resilience of the tensor power spectrum}  \label{resil}

Working in units $\tilde c_T =1$ the term involving the extrinsic curvature in the EFT action \cref{action} vanishes. The action of the tensor modes is in canonical form (at least to this order in the EFT expansion), and we can refer to this as the Einstein frame. The tensor power spectrum and spectral index then also take canonical form $\tilde{\Delta}_\gamma^2=\tilde{H}^2$ and $\tilde{n}_T-1 = -2\tilde{\eps}$, and in particular only depend on the Hubble parameter in the Einstein frame.

In \cite{Creminelli} this observation has been referred to as the ``resilience" of the power spectrum: the scale of inflation can always be uniquely defined in terms of the Einstein frame Hubble constant, and  the measurement of a blue tilt $\tilde n_T>1$ irrevocably implies a violation of the Null Energy Condition (NEC) in the Einstein frame $\partial_{\tilde{t}} \tilde{H}>0$. While we agree with these statements, viewed as a change of units this almost seems a tautology: the tensor spectrum depends in general on both the Hubble constant and the tensor speed, but working in units with the tensor speed set to unity, it only depends on the Hubble constant. One could equally change units such that $c_s\, \eps=1$ to argue for the resilience of the \emph{scalar} power spectrum: now the scalar amplitude only depends on the Hubble constant in this frame, and the observation of a blue scalar tilt indicates the breakdown of the NEC in this frame. The issue of course is that all of these statements are in terms of dimensionful quantities which are frame/unit dependent, and as such have no direct physical meaning. A more robust and frame invariant definition of the scale of inflation is the tensor power spectrum itself; if this becomes order one quantum gravity effects should be incorporated. We will discuss the frame-invariant formulation of the NEC in the next subsection.

Of course, while changing units does not change the physics, it can make the physics more transparent, and therefore the calculations easier.  Indeed, a huge advantage of the Einstein frame is that gravity is canonical, which simplifies computations of the tensor correlation functions such as the tensor bisprectum a lot, as was argued in \cite{Creminelli}.\footnote{Since the scalar modes are non-canonical in this frame, the mixed correlation functions are still non-standard, and there is in this sense a conservation of complexity.}
Also, the question to what extent inflation is a quantum gravity phenomenon is most transparently addressed in the Einstein frame. But these are all computational advantages, the physics is the same: independent of the units used, the correct computations will always yield the same answer for the (dimensionless and frame-invariant) observables, which contain all the physical information of the system.

\subsection{Further comments on recent literature}  \label{lit}

Although our main message ``the disformal transformation is just a change of units"  is de\-cei\-ving\-ly simple, it helps to understand results and clarify some confusion in recent literature.

To start with Refs.~\cite{Baumann, Baratella} have expressed worries that after the transformation to the Einstein frame $c_T \mapsto \tilde c_T=1$, the scalar speed of sound $c_s \mapsto \tilde c_s = c_s/c_T$ can exceed unity. Does  this apparant superluminality change the causal structure of the theory?  The answer is that there is no superluminality.  As follows from \cref{trafc}, changing units leaves the ratio of speeds constant
\be
\frac{c_s}{c_A} = \frac{\tilde c_s}{\tilde c_A}
\ee
with $c_A$ the light speed. Thus if the scalar speed is initially not larger than the speed of light, it also will not be superluminal after the transformation.  The confusion is that the scalar speed in the early universe is compared with the light speed {\it today}, which is not a useful comparison. Indeed, $c_s > 1$ means that the scalar speed exceeds the current speed of light $c = 3 \times 10^8$m/s.

In \cite{Creminelli}, for example, it was discussed that a blue tensor tilt does not automatically imply NEC conservation in non-Einstein frames. However, the statements $\partial_t H >0$ and $\partial_{\tilde t} \tilde H >0$ are in terms of dimensionful quantities.  They mean different things in different frames, and as such lose all their physical meaning.  It is only in the Einstein frame that the NEC is the basis of several gravitational theorems, such as that black holes cannot shrink and, in the context of inflation, that the tensor tilt cannot be blue shifted, and thus has a direct physical implication.  Using the notation defined in \cref{invariant}, where bars denote dimensionless quantities, the NEC can be written in frame invariant form
\be
\partial_{\bar t} \(\frac{\bar H}{\sqrt{\bar c_T}} \)> 0 \quad \Rightarrow \quad \partial_{\tilde t} \tilde H \big|_{\tilde c_T =1} > 0,
\label{NEC}
\ee
which reduces to the usual expression in the Einstein frame (right hand side).  Whatever units are used, it is only the above version of the NEC that has direct physical meaning; specifically, its violation implies a blue shifted tensor spectral index. Similar conclusions were found in \cite{NEC_new, NEC_new2}.

In \cite{Cai2} a model with time variations in $c_T$ is studied, which give rise to features in the tensor power spectrum. What happens after a disformal transformation that changes the tensor speed? Are the features still present in a frame  where $c_T=1$ is always constant? The authors of \cite{Cai2} resort to a numerical analysis to establish that also in the Einstein frame the tensor spectrum continues to exhibit the same features. Once one realizes that here one is investigating the effect of changing units on a dimensionless physical observable, these results come as no surprise.

Reference \cite{Sebastian} has investigated the structure of the acoustic peaks in the CMB for a non-canonical tensor speed. They thus assume $c_T \neq c_A$ during recombination, and the speeds enter the fluid equations describing the acoustic oscillations.\footnote {In this set-up, it is thus only after recombination that the tensor and light speed become equal (as they need be today). It is not clear whether such a scenario is in agreement with standard cosmology. Conversely in a set-up where $c_T $ and $ c_A$ become equal at the end of inflation, the CMB only depends on the power spectra at horizon exit, which are frame-independent.}  They consider a theory with canonical speeds for scalar modes and all matter fluids (baryons, DM, light)  $c_X=1$  and different values for the  tensor speed $c_T$. Analytical computations and numerical simulations together show that after a disformal transformation that sets $\tilde c_T=1$, one ends up with CMB peak structures corresponding to different matter speeds $\tilde c_X$. Changing units such that $c_T\mapsto 1$ implies that $c_{X} \mapsto \frac{1}{c_T}$, which directly explains the findings of \cite{Sebastian}. Note that different ratios $c_T/c_X$ correspond to different physical theories, and thus gives rise to different values for physical observables (as extracted from the CMB).  The reason is that this ratio enters the Einstein equations relating the (perturbations in) the spacetime curvature to the matter content in the universe, which  is non-canonical.  The ratio of speeds itself is of course invariant under a change of units.

As discussed in \cref{uchange} the equivalence of a disformal transformation and a change of units holds for the
general disformal transformation \cref{disformal}, and thus also to the special case of a conformal transformation.  There is a large literature on the equivalence of the Einstein and Jordan frame and the possible frame dependence of cosmological observables, see for example  \cite{framdep1, framdep2, framdep3, framdep4, framdep5}.
Realizing that the transformation between the two frames is nothing else than a (field-dependent) rescaling of the units, it is clear that this cannot affect the physics and that no observable can depend on it. Just as for the disformal transformation, much of the confusion arises because statements are made in terms of dimensionful variables. For example, comparing perturbations defined on time slices of equal energy between frames does not seem to be useful \cite{White}, as energy is not an invariant quantity; what can be compared are physical observables (or more general, frame-invariant quantities).

\section{Conclusions}  \label{con}


The disformal transformation is nothing else than a rescaling of length and time intervals, which can equi\-va\-lently be obtained from a unit change. This statement also holds for a time-dependent disformal transformation, and is valid up to all orders in perturbation theory.

Even if its impact on the actual physics is zero, as physical information comes exclusively from dimensionless and frame invariant quantities, changing units (or: performing a disformal transformation) can simplify computations considerably. Physical results are frame invariant, but in many cases the difficulty in extracting them from the theory is not. In particular, in the Einstein frame the graviton propagator is canonical. The price to pay is that one has to deal with a non-canonical scalar speed of sound, but that has been studied in the literature at great length. In this way, changing units (performing a disformal transformation) may enable one to extract the theory's physical contents in a faster and/or more transparent way.


Motivated by the amount of frame-related confusion found in the literature, we hope that the interpretation of the abstract disformal transformation as a change of units can provide some useful physical intuition.

\begin{appendix}

  \section{Invariant action}
  \label{invariant}

 In this appendix we rewrite the quadratic action and power spectrum in terms of dimensionless quantities, which are manifestly invariant under the time-dependent disformal transformation \cref{disformal}. The results can be straighforwardly be generalized to the action at all orders. As discussed in the main text we have to be careful with quantities that involve time-derivatives such as the Hubble constant.  Our approach is similar to the frame invariant constructions for a pure conformal transformation proposed in \cite{catena,volpe}.  In this formalism different units can be implemented by a different choice of length and time.  For mass units we use the Planck mass, which does not transform under the disformal transformation \cref{betatra} (i.e. we set $\mpl =1$).\footnote{To extend the formalism to the general disformal transformation \cref{disformal}, we would also need to include a unit of mass $M$.}

The dimensionless length and time  are defined as in \cref{equal}
\be
\bar a \dd \bar x = \frac{ a \dd  x}{L} = \frac{ \tilde a \dd  \tilde
  x}{ \tilde L}, \qquad
d\bar t = \frac{dt}{T} = \frac{d\tilde t}{\tilde T}.
\label{scaleLold}
\ee
The dimensionless velocity, Hubble parameter, and first slow roll parameter are 
\be
\bar c_i = c_i \frac{T}{L} = \tilde c_i \frac{\tilde T}{\tilde L},
\qquad
\bar H 
= \partial_{\bar t} \ln( \bar a \dd \bar x),
\qquad
\bar \eps = \frac{\partial_{\bar t} \bar H}{\bar H^2} .
\label{cT_dimold}
\ee
The dimensionless tensor and scalar actions are of the form
\begin{align}
\bar S_{\gamma} &= 
\int \dd \bar t \dd^3 \bar x\, \frac{1 }{8}\
  \bar a^3 
  (\bar c_T)^{-2} \[
\(\partial_{\bar t} \gamma_{ij}\)^2 - \bar c_T^2
  \frac{(\partial_{\bar k} \gamma_{ij})^2}{ \bar a^2} \]
\nn \\
\bar S_{\zeta} & =\int \dd \bar t \dd^3 \bar x
  \bar a^3
 (\bar c_s)^{-2} \bar \eps \[
\(\partial_{\bar t} \zeta\)^2 - \bar c_s^2
  \frac{(\partial_{\bar k} \zeta)^2}{ \bar a^2} \]
\end{align}
which leads to the power spectra
\be
\Delta_\gamma 
=  \frac{2 }{\pi^2 }\frac{ \bar H^2}{\bar c_T } , \qquad
\Delta_\gamma 
=\frac{ 1}{8\pi^2} \frac{ \bar H^2}{\bar c_s \bar \eps }.
\label{Pdimold}
\ee
The action and power spectrum are now manifestly invariant under the disformal transformation.  Restoring the dimensions, the action and power spectrum can be written in either the untilde or tilde variables, which retrieves the results before and after the transformation.  It is clear that this is equivalent to a change of units.

%
%
%

\end{appendix}

\end{document}